\documentclass[]{spie}  
\usepackage[]{graphicx}

\title{Superconducting metamaterials and qubits} 

\author{B.L.T. Plourde\supit{a}, Haozhi Wang\supit{a}, Francisco Rouxinol\supit{a},  M.D. LaHaye\supit{a}
\skiplinehalf
\supit{a}Department of Physics, Syracuse University, Syracuse, NY 13244-1130, USA
}

\authorinfo{Further author information: (Send correspondence to B.L.T.P.)\\B.L.T.P.: E-mail: bplourde@syr.edu}

\begin{document} 
  \maketitle 

\begin{abstract}
Superconducting thin-film metamaterial resonators can provide a dense microwave mode spectrum with potential applications in quantum information science. We report on the fabrication and low-temperature measurement of metamaterial transmission-line resonators patterned from Al thin films. We also describe multiple approaches for numerical simulations of the microwave properties of these structures, along with comparisons with the measured transmission spectra. The ability to predict the mode spectrum based on the chip layout provides a path towards future designs integrating metamaterial resonators with superconducting qubits.
\end{abstract}



\section{INTRODUCTION}
\label{sec:intro}  

Metamaterials having both a negative permeability and permittivity, and thus a negative index of refraction with left-handed transmission properties, were first described several decades ago \cite{Veselago1968}. More recently there have been numerous investigations of a variety of counterintuitive optical properties in these systems, including cloaking \cite{Alu08} and superlensing \cite{Pendry2000}. Research in this direction is closely connected to the field of photonic band-gap engineering \cite{Busch2007}.

In the microwave regime, left-handed transmission lines have been proposed and studied with one- and two-dimensional arrays of room-temperature lumped-element components \cite{Eleftheriades2002,Iyer09}. There have also been implementations of one-dimensional metamaterial microwave transmission lines with high-temperature superconducting films \cite{Wang2006}.

Since 1999, there has been substantial progress in the field of quantum coherent superconducting circuits, or qubits \cite{Clarke2008}. Many superconducting qubit implementations involve couplings between the qubits and superconducting microwave resonator circuits. In this architecture, referred to as circuit quantum electrodynamics, or cQED \cite{Wallraff2004, Blais2004, Schoelkopf2008}, the qubit behaves as an artificial atom that can couple to photons in the microwave resonant cavity, analogous to atomic QED with natural atoms in microwave or optical cavities \cite{Raimond2001}. In cQED, the microwave resonators are typically either distributed coplanar waveguide cavities patterned from superconducting thin films \cite{Goppl08} or three-dimensional waveguide cavities \cite{Paik11, Rigetti12}.

A recent theoretical work suggested the possibility of implementing one-dimensional superconducting metamaterial resonators in the microwave regime with high quality factors for coupling to superconducting qubits \cite{Egger2013}. Such a system could allow for the generation of large-scale entanglement between multiple photon modes in the metamaterial and could have potential applications in the emerging field of quantum simulation.

Here we report on the fabrication of thin-film superconducting metamaterial resonators and test oscillators along with microwave measurements at millikelvin temperatures. We also present numerical simulations of the metamaterial circuits and compare the simulated and measured spectra. The metamaterial architectures that we present here are compatible with future circuit designs for exploring the coupling to superconducting qubits.

\section{Design and Fabrication} 




\subsection{Metamaterials with microwave circuit components} 
\label{sec:metamaterials}

\begin{figure}
   \begin{center}
   \begin{tabular}{c}
   \includegraphics{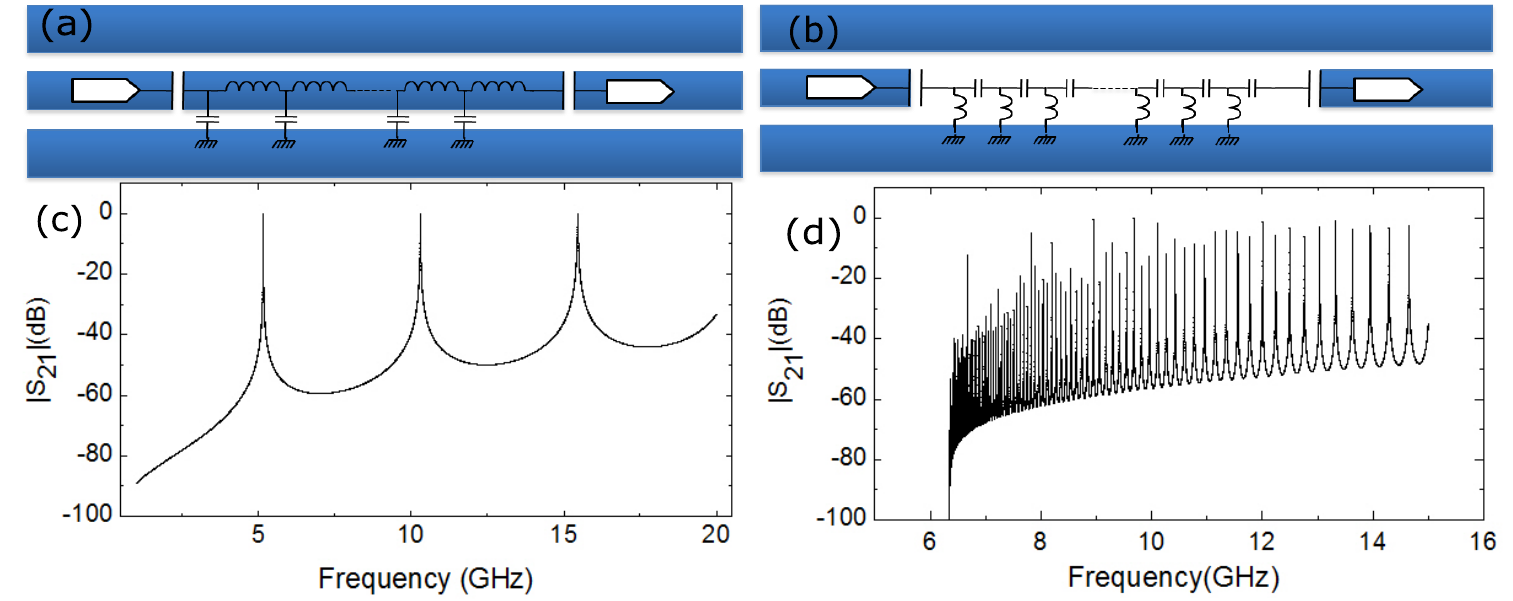}
   \end{tabular}
   \end{center}
   \caption[LHL-schem] 
   { \label{fig:circuit-schem} 
Circuit schematics for (a) right-handed and (b) left-handed lumped-element lines. $S_{21}$ spectra simulated in AWR Microwave Office for (c) right-handed and (d) left-handed transmission-line resonators.}
   \end{figure}

Although typical planar cQED implementations involve distributed coplanar waveguide (CPW) resonators fabricated from superconducting thin films on silicon or sapphire substrates \cite{Goppl08}, a resonator could also be formed from a lumped-element transmission line, consisting of a one-dimensional chain of unit cells each with a series inductor $L_l$ and a capacitor $C_l$ to ground [Fig.~\ref{fig:circuit-schem}(a)] \cite{Pozar05}. Such a resonant circuit would exhibit a fundamental resonance at $f_0=1/2 N\sqrt{L_l C_l}$, where $N$ is the number of unit cells between the input/output coupling capacitors. Beyond the fundamental resonance, there are evenly spaced harmonics at integer multiples of $f_0$, as shown in a circuit simulation performed in AWR Microwave Office [Fig.~\ref{fig:circuit-schem}(c)]. The dispersion relation for such a circuit is an increasing function of the wavenumber, characteristic of right-handed transmission. 

As described in Refs~[5, 16] \cite{Eleftheriades2002, Egger2013}, by interchanging the positions of the capacitors and inductors in the lumped-element right-handed transmission line resonator described above [Fig.~\ref{fig:circuit-schem}(b)], one obtains a circuit with dramatically different transmission properties. There is now a low-frequency infrared cut-off at $f_{IR}=1/4\pi\sqrt{L_l C_l}$, below which there is no transmission through the structure. For frequencies just above $f_{IR}$, there is a dense forest of resonances that get further apart for higher frequencies. The density of resonances for $f > f_{IR}$ increases as more unit cells are added to the transmission line. As with a conventional right-handed lumped-element transmission line, the impedance is still given by $Z_0 = \sqrt{L_l/C_l}$.

\subsection{Metamaterial design and fabrication} 

We design our metamaterial resonators with a target range of $f_{IR} \sim 5\,{\rm GHz}$ in order to place the densest portion of the metamaterial spectrum in the range of tunability for a typical superconducting transmon qubit \cite{Koch2007}. The use of an asymmetric transmon with two different sizes of junctions \cite{Strand13} would allow the IR cut-off for the metamaterial resonator to be located between the upper and lower high-coherence flux-insensitive sweetspots. Our target impedance for the metamaterial lines is $Z_0 = 50\,\Omega$, which, when combined with $f_{IR}=5\,{\rm GHz}$, results in $C_l = 300\,{\rm fF}$ and $L_l = 0.8\,{\rm nH}$. 

   \begin{figure}
   \begin{center}
   \begin{tabular}{c}
   \includegraphics{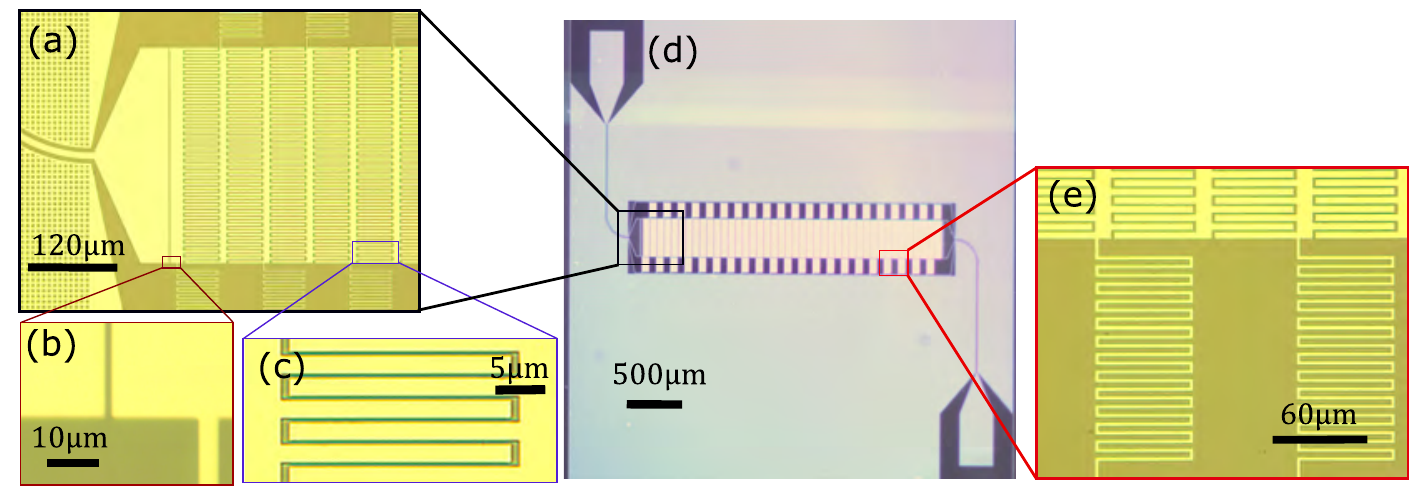}
   \end{tabular}
   \end{center}
   \caption[example] 
   { \label{fig:LHL-pics} 
Optical micrographs of type-A superconducting metamaterial transmission-line resonator: (a) close-up of first 5 unit cells, (b) coupling capacitor at one end of metamaterial, (c) close-up of interdigitated fingers in capacitor, (d) zoomed-out image of entire chip, (e) close-up image of two inductors.}
   \end{figure} 

We fabricate our metamaterial structures on a Si wafer with conventional photolithographic patterning using a DUV stepper followed by a lift-off process with electron-beam deposition of Al thin films, a common material in superconducting qubit circuits. The Al films are 90 nm thick. For our initial metamaterial structures we have used interdigitated capacitors, which allow for a straightforward single-layer fabrication process, although the capacitors occupy a relatively large area and their physical size limits the number of unit cells that we are able to fit on a chip. Our present capacitor design has fingers that are $4\,\mu{\rm m}$ wide, with a separation of $1\,\mu{\rm m}$ between adjacent fingers and an overlap length of $50\,\mu{\rm m}$ [Fig.~\ref{fig:LHL-pics}(a, c)]. For the inductors, we use a meander-line design with a linewidth of $2\,\mu{\rm m}$, gaps between adjacent lines of $3\,\mu{\rm m}$, with a length for each turn of the meander line of $60\,\mu{\rm m}$ [Fig.~\ref{fig:LHL-pics}(e)].

In order to explore the impact of the lumped-element parameter values on the resonance spectrum, we have fabricated two different versions of metamaterial resonators: type A and type B. In both cases, we have attempted to maintain the ratio of $L_l$ and $C_l$ for $Z_0\sim 50\,\Omega$ while varying their product to change $f_{IR}$. On the type A (B) structures, each capacitor has 32 (26) pairs of fingers and each inductor has 12 (7) turns of the meander line.

We have chosen the lumped-element parameters based on a variety of techniques for estimating the resulting capacitance and inductance values. For the capacitors, numerical simulations based on ANSYS Q3D yield a capacitance per finger pair of $8.35\,{\rm fF}$, corresponding to $C_l^A = 267\,{\rm fF}$ and $C_l^B = 217\,{\rm fF}$. 
For the inductors, 
from an analysis with Sonnet, we obtain inductance estimates $L_l^A=0.9\,{\rm nH}$ and $L_l^B=0.6\,{\rm nH}$. 
We note that these estimates do not account for kinetic inductance. However, for our relatively wide traces on the inductors and the short penetration depth of Al, we estimate the kinetic inductance contribution to be less than 5\% of the total inductance for each $L_l$.

Both metamaterial structures consist of 42 unit cells extending across the width of a $6\,{\rm mm}$-wide chip [Fig.~\ref{fig:LHL-pics}(d)] with gap coupling capacitors at either end of the metamaterial [Fig.~\ref{fig:LHL-pics}(b)]. For the type A(B) metamaterial resonators, the gap capacitor is $1\,(5)\,\mu{\rm m}$ wide, in order to target a coupling capacitance of $50\,(28)\,{\rm fF}$. Each chip has an Al ground plane surrounding the metamaterial structure and CPW leads for probing the microwave transmission. As is typically done in many cQED implementations, the ground plane contains a lattice of holes to avoid the trapping of Abrikosov vortices in regions of large microwave currents that could contribute excess loss \cite{Song2009}.

\section{Device measurements} 

\subsection{Measurements of test oscillators} 

\begin{figure}
   \begin{center}
   \begin{tabular}{c}
   \includegraphics{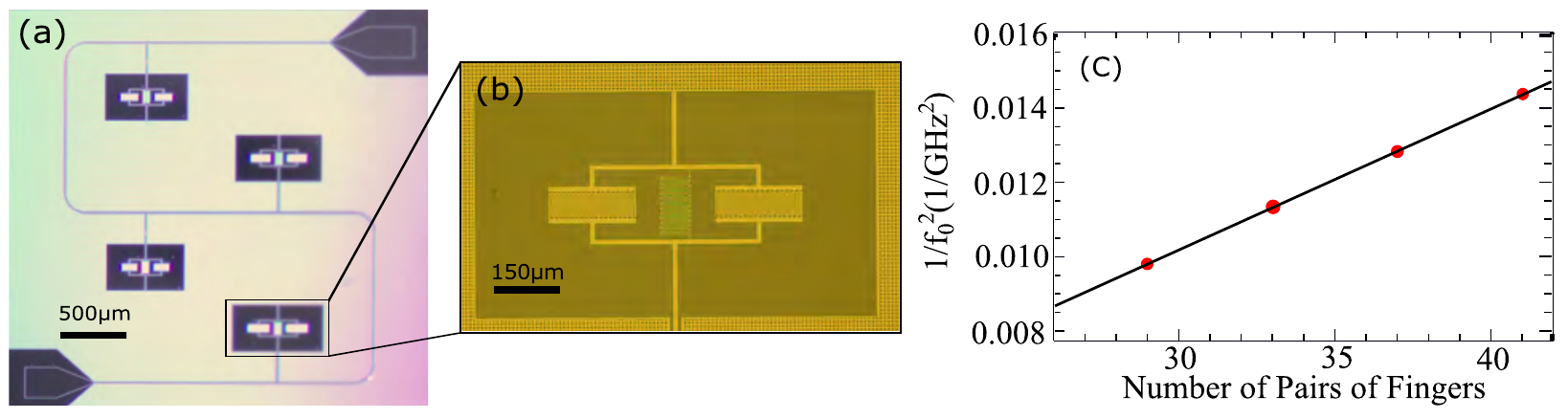}
   \end{tabular}
   \end{center}
   \caption[test-osc] 
   { \label{fig:test-osc-slope} 
Optical micrographs of test oscillator chip with parameters corresponding to type-A metamaterial resonator: (a) zoomed out; (b) close-up of single oscillator. (c) Plot of $1/f_0^2$ vs. number of finger pairs in capacitor of each test oscillator along with linear fit, as described in text.}
   \end{figure}

Before measuring the metamaterial resonators, we have designed and tested a series of lumped-element test oscillators fabricated with the same process and lumped-element parameters as in the metamaterials. The test oscillator chips each consist of four lumped-element LC oscillators, each one capacitively coupled to a CPW feedline. Each test oscillator has the same inductor as in a unit cell of metamaterial A or B and the capacitor has the same interdigitated finger parameters as in the corresponding metamaterial line, although the capacitor is split in two halves that are arranged in parallel on either side of the inductor. The total number of finger pairs in the capacitor for oscillator number 1-4 is 29, 33, 37, or 41, respectively, thus allowing us to study the variation in capacitance for different numbers of fingers [Fig.~\ref{fig:test-osc-slope}(a, b)].

We perform our measurements of the test oscillators on an adiabatic demagnetization refrigerator (ADR) at a temperature of $\sim 50\,{\rm mK}$. We probe the microwave transmission $S_{21}$ through the feedline with a vector network analyzer, sending microwave signals to the chip through a coaxial driveline with 53~dB of cold attenuation for thermalization. The transmission signal is amplified by a HEMT mounted on the $3\,{\rm K}$ plate of the ADR and again with a room-temperature amplifier. A cryogenic mu-metal can mounted on the $3\,{\rm K}$ plate surrounds the sample for magnetic shielding.

The resonance of each test oscillator results in a dip in $S_{21}$ and we fit the resonance trajectory in the complex plane with a standard form \cite{Pappas2011}, then extract the resonance frequency $f_0$ and quality factor $Q$ for each test oscillator. The internal loss values extracted from these fits are in the range of $10^{-5}$, consistent with loss due to two-level systems at interfaces and surfaces in thin-film superconducting circuits at low temperatures \cite{OConnell08,Pappas2011}. Upon obtaining $f_0$ for each test oscillator, we are able to plot $1/f_0^2$ vs. the number of finger pairs in the capacitor of each test oscillator [Fig.~\ref{fig:test-osc-slope}(c)]. The resulting linear variation should have a slope related to the product of the specific capacitance per finger pair and the common inductance value for each test oscillator. The slope in Fig.~\ref{fig:test-osc-slope}(c) combined with the specific capacitance per finger pair of $8.35\,{\rm fF}$ from the Q3D numerics, corresponds to an inductance of $L_l^A=1.15\,{\rm nH}$. A similar analysis for the type B metamaterial parameters results in $L_l^B=0.7\,{\rm nH}$. These inductance values are slightly larger than our Sonnet estimates from the previous section, which could be due in part to the lack of kinetic inductance in the Sonnet treatment. Also, since our analysis of the test oscillator data only gives us the product of the specific capacitance per finger pair and the inductance, it is possible that our capacitance values from Q3D could be off, which would result in a shift of the inductance values we extract. 

\subsection{Measurements of metamaterials} 

   \begin{figure}
   \begin{center}
   \begin{tabular}{c}
   \includegraphics{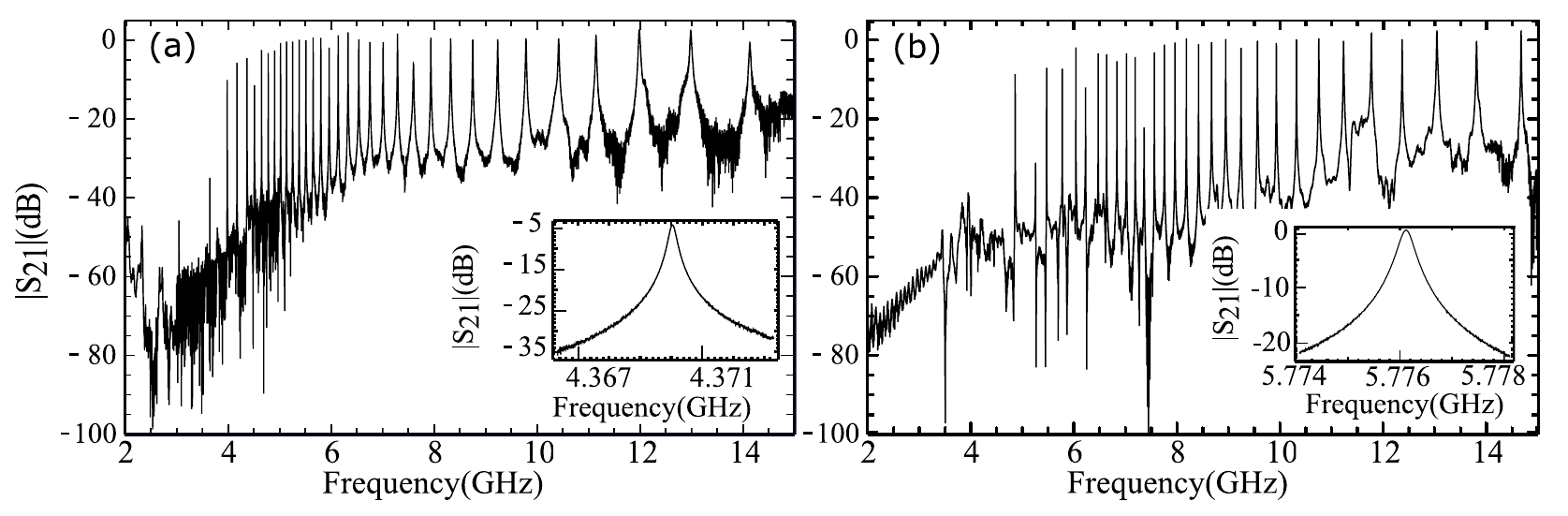}
   \end{tabular}
   \end{center}
   \caption[LHL-S21] 
   { \label{fig:LHL-S21} 
Measured $S_{21}(f)$ of metamaterial resonator (a) type A, (b) type B. Each curve is normalized with a separately measured baseline, as described in text. Insets show $S_{21}(f)$ for single peak near $f_{IR}$ for each metamaterial.}
   \end{figure} 

We cool our metamaterial resonators on the ADR to $\sim 50\,{\rm mK}$ with the same measurement configuration as described for the test oscillators in the previous section. Using the vector network analyzer, we probe $S_{21}(f)$ from $2-15\,{\rm GHz}$ through each metamaterial resonator, type A and B (Fig.~\ref{fig:LHL-S21}). In order to normalize the spectra, we measure the baseline transmission on our ADR below $100\,{\rm mK}$ on a separate cooldown with a superconducting feedline in place of the metamaterial resonator. Subtracting this baseline for each chip accounts for any sources of loss off of the chip, such as cables, connectors, and attenuators, as well as the gain from the amplifiers.

From the measured $S_{21}(f)$ spectra, there is a clear stop band at low frequencies, with low transmission below $\sim 4\,{\rm GHz}$ for type A and $\sim 5\,{\rm GHz}$ for type B, characteristic of an IR cut-off. Beyond this point, there are many resonance peaks that initially get closer together for increasing frequency, reaching a minimum peak spacing of $120\,(160)\,{\rm MHz}$ around $5\,(6)\,{\rm GHz}$ for type A (B) before spreading out to larger separations at higher frequencies. The 3~dB linewidths of the peaks just above $f_{IR}$ are $260\,(280)\,{\rm kHz}$ for the type A (B) structures, as shown in the insets to Fig.~\ref{fig:LHL-S21}.

\section{Simulations of circuit properties} 

\subsection{Simulations of metamaterials: numerical solutions to lumped-element circuit} 

\begin{figure}
   \begin{center}
   \begin{tabular}{c}
   \includegraphics{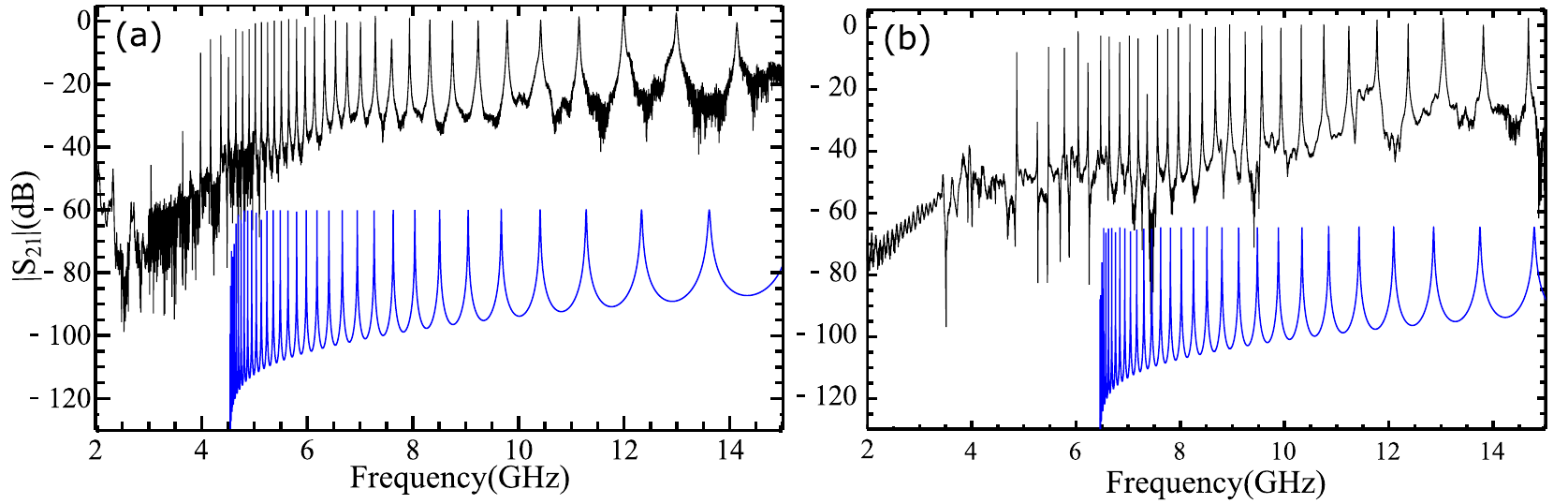}
   \end{tabular}
   \end{center}
   \caption[sims-AWR] 
   { \label{fig:sims-AWR} 
Simulated $S_{21}(f)$ using AWR Microwave Office for extracted circuit parameters as described in the text (blue line) and comparisons with measured spectra (black line) for metamaterial (a) type A, (b) type B. Simulated curves are offset by $-60\, (-65)\,{\rm dB}$ below the normalized measured spectra for type A (B).}
   \end{figure}

For our initial analysis of the measured $S_{21}(f)$ for the two metamaterial resonators, we compare these with numerical solutions to the lumped-element circuit model using AWR Microwave Office (Fig.~\ref{fig:sims-AWR}). In this approach, we take the values for $C_l$, $L_l$ and $C_c$ from the Q3D numerics and test oscillator analysis described in Sec. 2.2 and 3.1. The AWR simulations clearly indicate an IR cut-off, although this is roughly $1\,{\rm GHz}$ above the first peak in the measured spectra for both type A and type B. Also, the peak density in the AWR curves is highest just beyond the IR cut-off, whereas the measured $S_{21}$ spectra exhibit an increasing peak density that reaches a maximum $\sim 1-2\,{\rm GHz}$ beyond the frequency at which the first peak occurs. This difference between the measurements and the AWR simulations is likely due to deviations from the ideal lumped-element model with the standing-wave patterns being influenced by the distributed nature of the transmission in the physical structures. In order to explore this, we need another approach to the simulations that accounts for the actual circuit layouts.

\subsection{Simulations of metamaterials: finite-element approach} 

\begin{figure}
   \begin{center}
   \begin{tabular}{c}
 \includegraphics{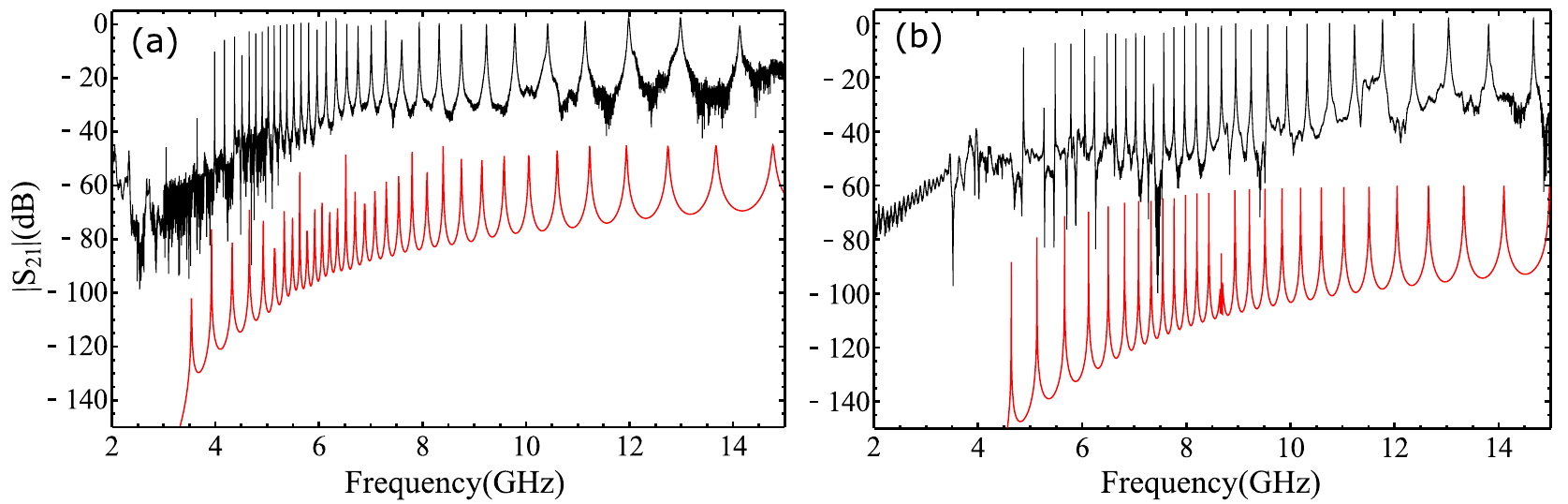}
   \end{tabular}
   \end{center}
   \caption[Sonnet] 
   { \label{fig:Sonnet} 
Simulated $S_{21}$ for circuit layout for metamaterial (a) type A and (b) type B using Sonnet (red line) as described in text; comparisons with measured spectra (black line). Simulated curves are offset by $-45\, (-60)\,{\rm dB}$ below the normalized measured spectra for type A (B).}
   \end{figure}

We simulate our physical metamaterial circuit layouts using the Sonnet software tool, which takes the CAD file for our circuit design as an input and solves Maxwell's equations to obtain the field distributions that can then be used to compute simulated $S_{21}(f)$ curves. From the comparison between the measured $S_{21}(f)$ and the Sonnet simulations in Fig.~\ref{fig:Sonnet}, we see that the Sonnet spectra exhibit a first peak that is within $\sim 0.5\,{\rm GHz}$ of the measured spectra. Also, the Sonnet spectra correctly capture the location of the maxima in the peak density being somewhat beyond the frequency of the first peak, in contrast to the lumped-element simulations from the previous section. The remaining deviations between the measured $S_{21}(f)$ and the Sonnet spectra could be due to various effects, including kinetic inductance in the Al traces, disorder in the values of $C_l$ and $L_l$ between the different unit cells, and coupling to chip modes in the Si substrate.

\section{Conclusions} 

\subsection{Conclusions} 
We have demonstrated that superconducting metamaterial resonators fabricated from Al thin films with interdigitated capacitors and meander-line inductors can exhibit a transmission spectrum with an IR cut-off and a dense mode spectrum of high-Q peaks. By varying the capacitor and inductor values in the metamaterial, we are able to modify the spectrum in a predictable way. Numerical simulations of the metamaterial structures agree with the measured transmission spectra reasonably well. The ability to predict the mode spectrum for a metamaterial transmission-line resonator from the circuit design parameters is an important stepping stone towards the integration of similar structures with superconducting qubits for applications in quantum information science, for example, following the strategy 
described by Egger \& Wilhelm \cite{Egger2013}.


\subsection{Acknowledgments} 
We acknowledge useful discussions with F.K. Wilhelm, D. Egger, and B. Taketani. This work was supported by the Army Research Office under Contract No. W911NF-14-1-0080. M.D.L. and F.R. also acknowledge support provided by the National Science Foundation under Grant No. DMR-1056423. Device fabrication was performed at the Cornell NanoScale Facility, a member of the National Nanotechnology Infrastructure Network, which is supported by the National Science Foundation (Grant ECS-0335765).

\bibliography{LHLPaper-Ref}   
\bibliographystyle{spiebib}   

\end{document}